\documentclass[preprint, preprintnumbers, amsmath ,amssymb, letterpaper, superscriptaddress]{revtex4}

\usepackage{amsmath}

\usepackage{graphicx}
\usepackage{dcolumn}
\usepackage{bm}

\usepackage{subfigure}

\preprint{}

\begin{document}

\title{\bf Linear Beam Stability in Periodic Focusing Systems: Krein Signature and Band Structure}

\author{Moses Chung} \email{mchung@unist.ac.kr}
\author{Yoo-Lim Cheon}
\affiliation{Department of Physics, Ulsan National Institute of Science and Technology, Ulsan 44919, Korea}
\author{Hong Qin}
\affiliation{Plasma Physics Laboratory, Princeton University, Princeton, New Jersey 08543}

\date{\today}

\begin{abstract}
The general question of how a beam becomes unstable has been one of the fundamental research topics among beam and accelerator physicists for several decades.
In this study, we revisited the general problem of linear beam stability in periodic focusing systems by applying
the concepts of Krein signature and band structure.
We numerically calculated the eigenvalues and other associated characteristics of one-period maps, and discussed
the stability properties of single-particle motions with skew quadrupoles and envelope perturbations in high-intensity beams
on an equal footing.

In particular, an application of the Krein theory to envelope instability analysis was newly attempted in this study.

The appearance of instabilities is interpreted as the result of the collision between eigenmodes of opposite Krein signatures
and the formation of a band gap.
\end{abstract}

\maketitle

\section{Introduction}
The general question of how a beam becomes unstable remains the central theme for accelerator physics.
Many fundamental stability problems in accelerator physics are described by
(homogenous) linear differential equations with periodic coefficients \cite{Yakubovich.book, Hong.MathPhys}.
The periodic coefficients are associated with the periodicity of the accelerators.
For example, circular machines have inherent periodicity of their circumferences,
and long linear accelerators or transport lines are
equipped with an arrangement of magnets with a repetitive sequence of identical modules \cite{Davidson.book, Wolski.book}.
Certainly, not only the accelerator physics, but also many other disciplines in science and engineering
have to deal with linear differential equations with periodic coefficients (see, for example, Ref. \cite{Yakubovich.book}).

For a simple system with one degree of freedom,
a linear differential equation with
a periodic coefficient takes
the form of a harmonic oscillator with a periodic spring constant expressed as \cite{Hong.MathPhys}
\begin{equation}
\frac{d^2 x(s)}{d s^2} + \kappa (s) x(s) = 0,
\end{equation}
where $\kappa(s) = \kappa(s + L)$ with periodicity $L$, and $s$ is the path length that plays the role of a time-like variable.
Mathematically, this equation is also known as Hill's equation.
In the context of uncoupled charged particle dynamics,
Courant and Snyder \cite{CS} analyzed the stability of
an alternating-gradient synchrotron using a $2\times2$ transfer matrix
in an elegant way.
When there exits some coupling between two degrees of freedom, the transfer matrix needs to be $4\times4$,
and the stability analysis becomes more complicated in terms of physics.

In circular and linear accelerators,
systems with two degrees of freedom usually appear in two ways:
single-particle motions around the reference orbit with a state vector ${\bf z} = (x, y, x', y')^T$ and
evolution of perturbations around matched beam envelopes with a state vector ${\bf z} = (\delta a, \delta b, \delta a', \delta b')^T$ \cite{Li.PRSTAB}.
For single-particle motions, skew quadrupole or solenoidal components in the beam line couple the $x$ and $y$ motions, whereas
for beam envelope perturbations, linear space-charge forces \cite{KV} are the sources of coupling.
For more detailed and complete stability analyses (e.g., in the presence of both external coupling and space-charge force),
all the second-order moments \cite{Yuan.PRSTAB} or collective motions \cite{Li.PRSTAB} may be taken into account.
When three-dimensional effects are important, the longitudinal phase space should be included as well, similar to Refs. \cite{Ji.PRSTAB1,Ji.PRSTAB2}.
However, to introduce the concept of Krein signature and band structure in this study,
we only focused on two basic cases where
the state variables evolve according to
\begin{equation}
\frac{d {\bf z}}{d s} = K(s) {\bf z}, ~~ K(s) = K(s + L),
\label{K}
\end{equation}
and the solution is provided in terms of a $4\times4$ transfer matrix $M(s)$ as \cite{Lund.PRAB2004}
\begin{equation}
{\bf z}(s) = M(s) {\bf z}_0,
\end{equation}
where $K(s)$ is a $4\times4$ focusing matrix and ${\bf z}_0 = {\bf z}(s=0)$ is the initial state vector.
Equation (\ref{K}) can be regarded as a matrix version of the Hill's equation.

A system of linear differential equations with periodic coefficients of two degrees of freedom
can be cast into a Hamiltonian system.
Particularly, for the charged particle dynamics considered in this study, the Hamiltonian is real.
The linear stability of this system is then determined by the one-lattice period map $M(L)$ (or one-turn map in a circular machine).
Because of the real Hamiltonian nature of the system, $M(L)$ is a real symplectic matrix.
It is well-established in Refs. \cite{CS, Dragt, Struckmeier, Reiser.book, Conte_MacKay, Lund.PRAB2004} that
the eigenvalues of $M(L)$ define the stability properties of the system.
The four eigenvalues $\lambda_n$ should be either complex conjugate pairs or real numbers.
Simultaneously,
from the symplectic condition, we obtain $\det[ M(L) ] = \Pi_{n=1}^4 \lambda_n = 1$.
The eigenvalues can be put into polar form:
\begin{equation}
\lambda_n = \gamma_n \exp( i \sigma_n),
\end{equation}
where $\gamma_n = | \lambda_n |$ is the growth factor of $n$-th mode per one lattice period
and $\sigma_n$ is the phase advance of that mode per one lattice period \cite{Lund.PRAB2004}.
If all the eigenvalues lie on the unit circle (i.e., $|\lambda_n| = 1$) and they are distinct,
the system is known to be stable (or more precisely, strongly stable \cite{Yakubovich.book}).
As noted earlier, several studies on beam stability analysis based on eigenvalues have been conducted.
There is an excellent review
on this subject by Lund and Bukh \cite{Lund.PRAB2004}.

As the system parameter varies from a stable equilibrium, the Hamiltonian and its corresponding $M(L)$ are deformed
and the eigenvalues move along the unit circle \cite{Yakubovich.book}.
It is well-known that a necessary and sufficient condition for the onset of instability is a collision
between eigenvalues of different types (i.e., with opposite Krein signatures defined later in this paper) \cite{Krein, Gel'fand, Moser, Yakubovich.book, Zhang.PoP}.
After the collision, the eigenvalues generally move off the unit circle as we further change the system parameter.

The concept of Krein collision has been adopted to accelerator physics by several authors, for example in Ref. \cite{Forest}.
More recently in Refs. \cite{GCS.PRSTAB,Spectral_Structural},
Krein signature was introduced to clarify the stability properties of single-particle dynamics.
An application of the Krein signature to envelope instability is newly attempted in this study.

In applied mathematics \cite{Yakubovich.book, Arnold.book} and nonliner dynamics \cite{nonlinear1, nonlinear2},
a one-lattice period map $M(L)$ is also known as monodromy matrix or Floquet matrix.
The dynamics of its eigenvalues (often called Floquet multipliers, characteristic multipliers, or simply, multipliers) resulting from changing the system parameter
has been actively investigated from various perspectives.
For example, in Refs. \cite{Aubry1,Aubry2,band1,band2,band3}, a band structure analysis
was adopted to better understand the occurrence of instabilities through Krein collisions in discrete breathers.
Recently, in the plasma physics community, the results of Krein analysis have been applied to
a complex G-Hamiltonian system (i.e., a complex generalization of the usual real Hamiltonian) \cite{Yakubovich.book}.
It was found that the physical meaning of the Krein signature is the sign of the action for the eigenmode,
and the only route for instability is through the resonance between the positive- and negative-action modes \cite{Zhang.PoP}.

In this study, we revisited the stability analysis based on the eigenvalues of a
one-period map in accelerator physics
from the perspective of recent findings on the Krein collision.
In Sec. \ref{II}, we introduce two model systems described by linear differential equations with periodic coefficients.
These systems are often found in accelerator physics.
In Sec. \ref{III}, the Hamiltonian formulation and eigenvalue analysis of a
one-period map are reviewed.
We then discuss recent advances in Krein theory and band structure analysis in Secs. \ref{IV} and \ref{V}, respectively.
Section \ref{VI} is devoted to numerical examples illustrating the properties of the Krein collision within the context of
beam stability.
Finally, we draw conclusions in Sec. \ref{VII} with a discussion on future research directions.

\section{Model Systems} \label{II}

\begin{figure}
  \centering
  \includegraphics[width=10cm]{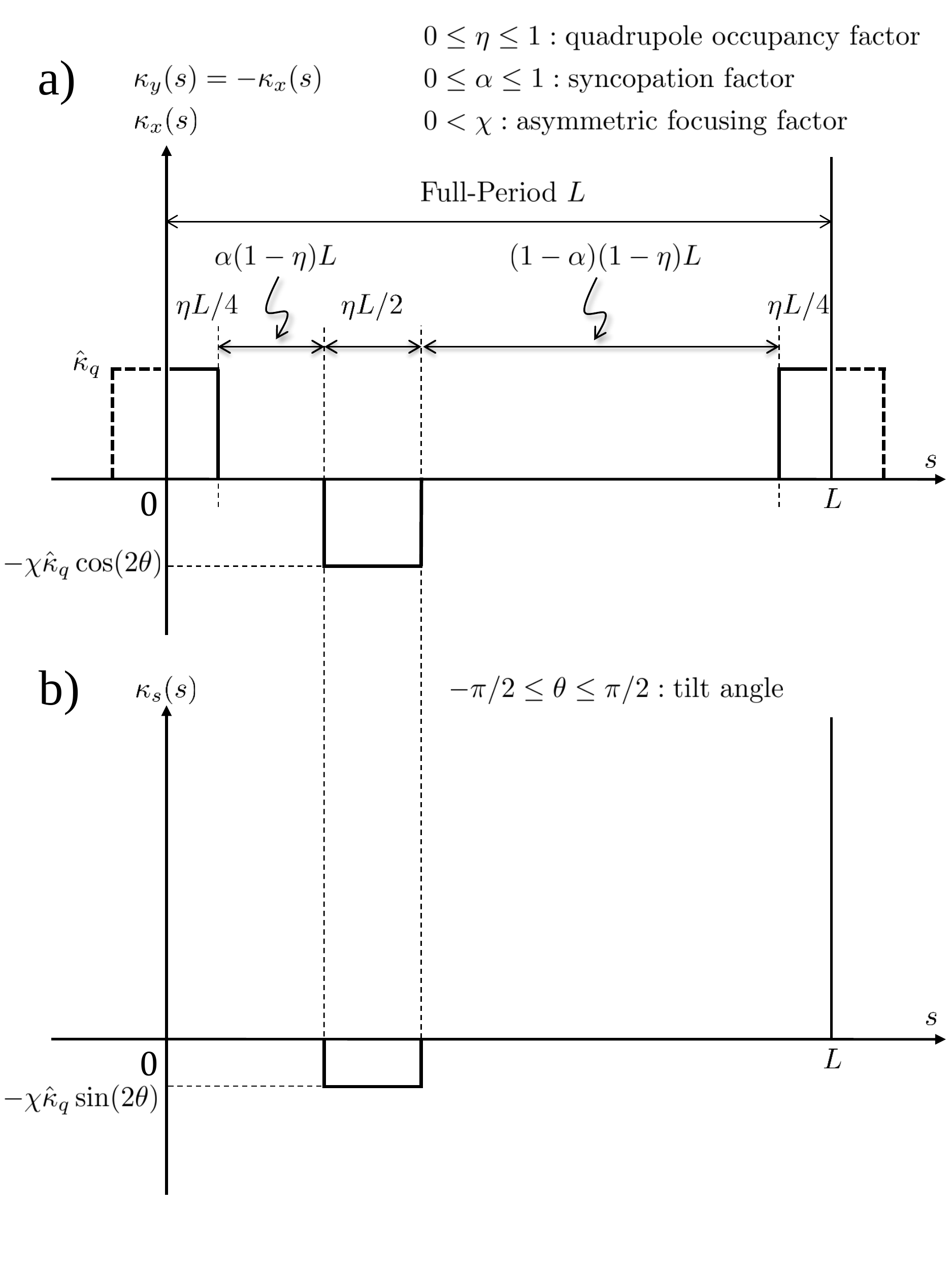}\\
  \caption{
  (a) Linear focusing coefficient $\kappa_x(s) = - \kappa_y(s)$ of the periodic quadrupole doublet lattice \cite{Lund.PRAB2004} with a different focusing strength for the middle magnet.
  (b) Smooth-focusing (or continuous-focusing) model with a skew quadrupole component $\kappa_{sq}(s)$.}
  \label{Fig1}
\end{figure}

We consider a piece-wise constant quadrupole doublet lattice depicted in Fig. \ref{Fig1}(a) \cite{Lund.PRAB2004}.
The focusing strength within the quadrupole's axial length ($\eta L/2$) can be expressed as
\begin{equation}
\kappa_q
= \frac{1}{[B \rho]} \left( \frac{\partial B_x^q}{\partial y}\right)_{(0,0)}
= \frac{1}{[B \rho]} \left( \frac{\partial B_y^q}{\partial x}\right)_{(0,0)}
= \frac{q_b B_q'}{p_0},
\end{equation}
where $p_0$  is a fixed reference momentum, $q_b$ is the charge of a beam particle, and $[B \rho] = p_0/q_b$ is the magnetic rigidity.
Here, the quadrupole magnetic field near the beam axis $(x, y) = (0, 0)$ is approximated by ${\bf B}_q = B_x^q \hat{x} + B_y^q \hat{y}$ to leading order \cite{Davidson.book}.
A positive quadrupole strength $\kappa_q$ in the $x$-plane inherently assumes that
in the $y$-plane the quadrupole strength is negative with the same magnitude ($-\kappa_q$), and vice versa \cite{CASnote}.
Hence, we cannot create a quadrupole magnet with different focusing amplitudes in each plane.
Instead, by adjusting $\chi$ (asymmetric focusing factor) of the adjacent magnet,
it is possible to make the average focusing effects in two planes different.
In Fig. \ref{Fig1}(a), $L$ is the lattice period, $\hat{\kappa}_q = |\kappa_q|$ is the constant amplitude of the quadrupole,
$\eta$ is the occupancy factor, and $\alpha$ is the syncopation factor \cite{Lund.PRAB2004}.
For $\alpha= 1/2$, we have a symmetric FODO (Focusing-Off-Defocusing-Off) lattice.

To examine the effects of coupling, we also consider a skew quadrupole component, as depicted in Fig. \ref{Fig1}(b).
For the quadrupole doublet lattice, the phase advance per period should be $\sigma <  180^\circ$ for single-particle stability.
In this case, it is not possible to meet the condition for sum resonances, which is $\sigma_x + \sigma_y = 2 \pi \times \mbox{integer}$ \cite{Conte_MacKay, Wolski.book}.
Hence, to demonstrate the onset of sum resonances in our model system, we adopt the smooth-focusing lattice represented by Fig. \ref{Fig1}(b).
In the smooth-focusing model, the phase advance can be adjusted to an arbitrary value.
Furthermore, to allow tunes (i.e., $\nu = \sigma/ 2 \pi$) to be different in two directions,
we again introduce the asymmetric focusing factor $\chi$.
In the numerical examples for single-particle motions, we fix $\kappa_x = \kappa_{sf}$ and vary $\kappa_y$ by $\chi \kappa_{sf}$.

\subsection{Single-particle motions with skew quadrupoles}\label{Sec_single}
When skew quadrupole components are present in addition to the standard quadrupoles,
the transverse dynamics in the $x-$ and $y-$dimensions are coupled.
Such couplings between two dimensions can be introduced either intentionally \cite{G_KV3},
or as a result of misalignment of the quadrupole magnet \cite{Wolski.book, G_KV}.
The single-particle linear dynamics with a skew quadrupole is described by
\begin{eqnarray}
\frac{d x}{d s}   &=& x', \\
\frac{d y}{d s}   &=& y', \\
\frac{d x'}{d s}  &=& - \kappa_x x -  \kappa_{sq} y, \\
\frac{d y'}{d s}  &=& - \kappa_{sq} x - \kappa_y y.
\end{eqnarray}
In terms of matrix form, we have
\begin{equation}
\frac{d {\bf z}(s)}{ds} = K(s) {\bf z}(s),
\label{perturbation1}
\end{equation}
where the components of ${\bf z} = (x, y, x', y')^T$ are the phase-space coordinates
of the particle motions and the matrix $K(s)$ is given by
\begin{equation}
K(s) =
\left(
  \begin{array}{cccc}
    0       & 0 & 1 & 0 \\
    0       & 0 & 0 & 1 \\
    -\kappa_x    & -\kappa_{sq} & 0 & 0 \\
    -\kappa_{sq} & -\kappa_y    & 0 & 0 \\
  \end{array}
\right)
=
\left(
                \begin{array}{cc}
                  0        & I \\
                  -\kappa_m & 0 \\
                \end{array}
\right),
\end{equation}
where $\kappa_m$ is a $2\times2$ symmetric matrix.
Note that the condition $K(s) = K(s+L)$ is fulfilled for the smooth focusing lattice shown in Fig. \ref{Fig1}(b).
The focusing matrix $\kappa_m$ is not constant because $\kappa_{sq}(s)$ is non-zero only for a certain range of the lattice.

\subsection{Envelope perturbations in high-intensity beams} \label{Sec_envelope}
The analysis of small-amplitude perturbations around matched beam envelopes
has been used as a basic theoretical tool to characterize high-intensity beam transport \cite{Struckmeier,Reiser.book, Lund.PRAB2004}.
In the periodic focusing quadrupole field,
the evolutions of the $x$- and $y$-direction envelopes of the Kapchinskij-Vladimirskij (KV) \cite{KV} distribution beam, $a(s)$ and $b(s)$,
are described by \cite{Davidson.book}
\begin{equation}
a'' (s) + \kappa_x a(s) - \frac{K_b}{a(s) + b(s)} - \frac{\epsilon_x^2}{a^3(s)}  =0,
\label{a}
\end{equation}
\begin{equation}
b'' (s) + \kappa_y b(s) - \frac{K_b}{a(s) + b(s)} - \frac{\epsilon_y^2}{b^3(s)}  =0.
\label{b}
\end{equation}
The envelope equations (\ref{a}) and (\ref{b}) represent a system of
two nonlinear, second-order coupled differential equations.
Given that the analytical solutions are not available in general,
the envelope equations (\ref{a}) and (\ref{b}) should be solved numerically for prescribed initial conditions
${\bf z}(0) = \left[ \delta a (0), \delta b (0), \delta a' (0), \delta b' (0) \right]^T$.
The dimensionless parameter $K_b$ is the self-field perveance defined either in terms of line density $N_b$, bunch current $I_b$, or line charge density $\lambda_b$ as
\begin{equation}
K_b
= \frac{1}{4 \pi \epsilon_0} \frac{2 N_b q_b^2}{\gamma_0^2 \beta_0 c p_0}
= \frac{1}{2 \pi \epsilon_0} \frac{q_b I_b}{\gamma_0^2 v_0^2 p_0}
= \frac{1}{2 \pi \epsilon_0} \frac{q_b \lambda_b}{\gamma_0^2 \beta_0 c p_0},
\end{equation}
where $p_0 = \gamma_0 m_b \beta_0 c$ is a fixed reference momentum.
The total emittances (100\% or rms edge emittances) are given by
\begin{equation}
\epsilon_x = 4 \left[ \left< x^2 \right>\left< x'^2 \right> - \left< x x'\right>^2 \right]^{1/2},
\end{equation}
\begin{equation}
\epsilon_y = 4 \left[ \left< y^2 \right>\left< y'^2 \right> - \left< y y'\right>^2 \right]^{1/2}.
\end{equation}

To investigate the stability of perturbations around the matched beam envelopes,
we linearize the envelope equations (\ref{a}) and (\ref{b}) as follows:
\begin{equation}
a (s) = a_m (s) + \delta a (s),
\end{equation}
\begin{equation}
b (s) = b_m (s) + \delta b (s).
\end{equation}
Here, $a_m(s)$ and $b_m(s)$ are the periodic matched-beam solutions with
\begin{equation}
a_m (s) = a_m (s + L),~~ a_m' (s) = a_m' (s + L),
\end{equation}
\begin{equation}
b_m (s) = b_m (s + L),~~ b_m' (s) = b_m' (s + L),
\end{equation}
where $L$ is the lattice period.
The linearized perturbation equations are then given by
\begin{eqnarray}
\frac{d}{d s} \left( \delta a \right)  &=& \delta a', \\
\frac{d}{d s} \left( \delta b \right)  &=& \delta b', \\
\frac{d}{d s} \left( \delta a' \right) &=& - \kappa_x \delta a -  \frac{2 K_b}{ (a_m + b_m)^2 } ( \delta a + \delta b )  - \frac{3 \epsilon_x^2}{a_m^4} \delta a, \\
\frac{d}{d s} \left( \delta b' \right) &=& - \kappa_y \delta b -  \frac{2 K_b}{ (a_m + b_m)^2 } ( \delta a + \delta b )  - \frac{3 \epsilon_y^2}{b_m^4} \delta b.
\end{eqnarray}
In terms of matrix form, we obtain
\begin{equation}
\frac{d {\bf z}(s)}{ds} = K(s) {\bf z}(s),
\label{perturbation2}
\end{equation}
where the components of ${\bf z} = (\delta a, \delta b, \delta a', \delta b')^T$ are the phase-space coordinates
of the envelope perturbations and the matrix $K(s)$ is given by
\begin{equation}
K(s) =
\left(
  \begin{array}{cccc}
    0       & 0 & 1 & 0 \\
    0       & 0 & 0 & 1 \\
    -k_{xm} & - k_{0m} & 0 & 0 \\
    -k_{0m} & - k_{ym} & 0 & 0 \\
  \end{array}
\right)
=
\left(
                \begin{array}{cc}
                  0        & I \\
                 - \kappa_m & 0 \\
                \end{array}
\right),
\end{equation}
where $\kappa_m$ is a $2\times2$ symmetric matrix.
Here,
\begin{equation}
k_{xm} =\kappa_x + \frac{3 \epsilon_x^2}{a_m^4} + k_{0m},
\end{equation}
\begin{equation}
k_{ym} =\kappa_y + \frac{3 \epsilon_y^2}{b_m^4} + k_{0m},
\end{equation}
\begin{equation}
k_{0m} =\frac{2 K_b}{(a_m +b_m)^2}.
\label{km}
\end{equation}
Note that the condition $K(s) = K(s+L)$ is fulfilled for periodic matched-beam solutions $a_m$ and $b_m$.

\section{Hamiltonian Formulation and Eigenvalue Analysis} \label{III}
For both single-particle motions with skew quadrupoles (Sec. \ref{Sec_single}) and envelope perturbations in high-intensity beams (Sec. \ref{Sec_envelope}),
the dynamics of the state vector $\bf z$ is described by
\begin{equation}
{\bf z}' = J A {\bf z}.
\label{z'}
\end{equation}
Here, $J$ is a $4 \times 4$ unit symplectic matrix,
\begin{equation}
J = \left(
      \begin{array}{cc}
        0 & I \\
        -I & 0 \\
      \end{array}
    \right),
\end{equation}
and $I$ is the unit matrix. Note that $J^T = J^{-1} = - J$.
The  matrix $A$ is then
\begin{equation}
A = J^{-1} K = \left(
                \begin{array}{cc}
                  \kappa_m & 0 \\
                  0 & I \\
                \end{array}
              \right).
\end{equation}
Given that $\kappa_m$ is symmetric, the matrix $A$ is also symmetric.
Even though one model is dealing with single-particle motions and the other model collective beam oscillations,
both systems in Secs. \ref{Sec_single} and \ref{Sec_envelope} are mathematically identical.

The corresponding Hamiltonian function is
\begin{equation}
H(z) = \frac{1}{2} {\bf z}^T A {\bf z}.
\end{equation}
The solution of this system can be expressed
as a symplectic linear map $M(s)$
\begin{equation}
{\bf z}(s) = M(s) {\bf z}_0,
\end{equation}
where ${\bf z}_0$ denotes arbitrary initial conditions at $s=0$ and $M(0)$ is a $4\times4$ identity matrix.
According to Floquet theory,
\begin{equation}
M(s + L) = M(s) M(L), ~~M(s + nL) = M(s) M(L)^n,
\end{equation}
where $n$ is an integer.
The linear stability of the system is then
determined by the eigenvalues of $M(L)$.
According to Ref. \cite{Lund.PRAB2004, Dragt}, the eigenvalues of $M(L)$
are determined uniquely regardless of the location of the initial position.
The eigenvalue equation, $\det[M(L) - \lambda I ] = 0$, does not change its form.
Hence, without loss of generality, we only consider the case in which the initial conditions are given at $s=0$.
Note that generally the eigenvectors are calculated differently when a different initial position is employed.

\section{Krein Signature} \label{IV}

Let $\psi$  be an eigenvector (or eigenmode) of a one-period transfer map $M(L)$.
The Krein product is then defined as
\begin{equation}
\left< \psi, \psi \right> \equiv \psi^\dagger (- i J) \psi,
\end{equation}
where $\psi^\dagger = (\psi^*)^T$ is the conjugate transpose of $\psi$.
Depending on the authors,
the sign of the product was chosen to be opposite by defining the inner product or $J$ differently.
This is just a matter of convention.
The sign of the Krein product for a given eigenmode with the eigenvalue $\lambda$ is called Krein signature and is defined as
\begin{equation}
\kappa(\lambda) = {\rm sgn} \left( \left< \psi, \psi \right> \right).
\end{equation}
According to this definition, the Krein signature can have $-1$, $0$, and $+1$.
When the eigenvalues stay on the unit circle, their Krein signatures are either $-1$ or $+1$.
If the eigenvalues move off the unit circle, their Krein signatures are assigned to be $0$.
Several important properties of the Krein product and its signature have been
discovered or re-discovered by many authors since the original work by Krein was published in 1950 \cite{Krein}.

\begin{enumerate}
\item The Krein product is a symplectic invariant.
For a symplectic transformation $M(s)$,
\begin{equation}
\left< M(s) \psi, M(s) \psi \right> = \psi^\dagger M(s)^T ( - i J) M(s) \psi = \psi^\dagger (-i J) \psi = \left< \psi, \psi \right>,
\end{equation}
where $M(s)^T J M(s) = J$.

\item A one-period transfer map $M(L)$ can be formally expressed as \cite{Hong.MathPhys}
\begin{equation}
M(L) = \exp( \bar{K} L) = \exp( J \bar{A} L),
\end{equation}
where $\bar{K} = J \bar{A}$ is a matrix representing the averaged effect of $K(s)$ in one period $L$.
The eigenvalue of $M(L)$ can be written as $\lambda = \exp(i k L )$ with $k$ being the wavenumber of the eigenmode.
An eigenvector of $M(L)$ is also an eigenvector of  $J \bar{A}$, i.e., $J \bar{A} \psi = i k  \psi$.
Therefore,
\begin{equation} \label{dimension}
\left< \psi, \psi \right> =  \psi^\dagger (-i J) \psi =  \frac{ \psi^\dagger \bar{A}  \psi }{k} =  2 \frac{\bar{H} ( \psi ) }{k}.
\end{equation}
Here, $\bar{H} ( \psi ) = \frac{1}{2} \psi^\dagger \bar{A}  \psi$ is the average energy (Hamiltonian) of the eigenmode.
It should be emphasized that the physical meaning of the Krein product is action,
i.e., the ratio between the energy and the frequency (or wavenumber) of the eigenmode, with neglecting an unimportant factor.
Given that we use the spatial coordinate $s$ as a time-like variable,
the dimension of the action in our case is [normalized energy]$\times$[length]
which is indeed the dimension of Eq. (\ref{dimension}).
Therefore, the algebraic meaning of the Krein signature is the sign of the action of an eigenmode \cite{Hong.MathPhys, Zhang.PoP}.
Some studies interpret the Krein signature as the sign of mass \cite{SIAM_News}
or the sign of energy \cite{Bridges}.

\item
If we let ${\bf u}(s)$ and ${\bf v}(s)$ be any two solutions of the linear equation (\ref{z'}), then
we have \cite{Bridges}
\begin{equation}
\frac{d}{ds} \left<{\bf u} , \bf v \right> = 0.
\end{equation}
This is one of the forms of action conservation. Indeed, this statement is equivalent to
the item 1, which says that the Krein product is a symplectic invariant.
In continuum mechanics, the physical quantity in item 2, $\bar{H} ( \psi )/ k$, is often called wave action or wave action density
and is used as `a conservable measure of the wave part of a motion' \cite{Wiki}.
It is well-known that
when a wave's frequency (or wavenumber in this study) increases (decreases), the wave gains (loses) energy to preserve wave action \cite{action}.

\item
When a system parameter changes in such a way that $[ \det( \bar{A} )]^{1/2}$ is increasing
(i.e., the average focusing effects of $A$ or $\kappa_m$ become stronger),
the Floquet multipliers (eigenvalues) of the positive Krein signature on the unit circle move
counterclockwise (i.e., in the sense of increasing the Floquet argument or the phase advance of the eigenmode) \cite{Bridges, Lectures_on_Dynamical_Systems}.
On the contrary,
the Floquet multipliers of the negative Krein signature behave oppositely, moving in the clockwise direction.
In this regard, only the phase advances of positive-signature eigenmodes have correct physical meanings.
For the two degrees of freedom,
we denote the phase advances of two positive-signature eigenmodes by $\sigma_1$ and $\sigma_2$.

\item (Krein-Gel'fand-Lidskii theorem)
According to the theory developed by Krein, Gel'fand, and Lidskii \cite{Krein, Gel'fand, Moser},
a linear Hamiltonian system is strongly stable if and only if
all the multipliers lie on the unit circle, and no multipliers of different signature collide \cite{Yakubovich.book}.
Here, the definition of strong stability is as follows:
a Hamiltonian system is said to be strongly stable if i) all the solutions of the equation are
bounded on $(-\infty, +\infty)$ and ii) this property is preserved by any small deformation of the Hamiltonian \cite{Yakubovich.book}.

\end{enumerate}

The detailed proofs of the above properties are given in Refs. \cite{Yakubovich.book, Hong.MathPhys, Bridges}.
As the parameters of a stable periodic-coefficient linear Hamiltonian system vary,
the multipliers move on the unit circle and may collide with each other.
A necessary and sufficient condition for the onset of instability is that
two multipliers with opposite Krein signatures collide \cite{Lectures_on_Dynamical_Systems, Zhang.PoP}.
This phenomenon is known as Krein collision \cite{Krein, Gel'fand, Moser}.
The underlying physical mechanism of the Krein collision is that the system
is destabilized when and only when a positive-action mode resonates with a negative-action mode \cite{Zhang.PoP}.

On special occasions, two multipliers with opposite Krein signatures may pass through each other without moving off the unit circle.
This happens when the Hamiltonian is deformed in a specific way as a certain system parameter varies.
Nonetheless, if we slightly change other system parameters at the Krein collision point,
we can always deform the Hamiltonian so that the multipliers eventually move off the unit circle.
Indeed, the Krein-Gel'fand-Lidskii theorem states that
in the neighborhood of a Krein collision with opposite signatures, there always exists an unstable periodic Hamiltonian \cite{Yakubovich.book}.

Therefore, by monitoring the Krein collisions together with the signatures of the involved eigenvalues,
regardless of whether the eigenvalues are moving off the unit circle immediately or not,
it is possible to effectively identify the parameter space for the onset of instabilities.

\section{Band Structure} \label{V}
Equations (\ref{perturbation1}) and (\ref{perturbation2}) can be put into a second-order coupled differential equation as
\begin{equation}
\xi'' + \kappa_m \xi = 0,
\label{xi1}
\end{equation}
where $\xi = (x, y)^T$ or $(\delta a, \delta b)^T$.
By introducing an eigenvalue $E$, it is possible to obtain an eigenvalue equation of Sturm-Liouville type as follows \cite{Aubry1}:
\begin{equation}
\xi'' + \kappa_m \xi = E \xi.
\label{xi2}
\end{equation}
The solutions of Eq. (\ref{xi1}) [or equivalently, Eqs. (\ref{perturbation1}) and (\ref{perturbation2})]
can be regarded as the eigenfunctions of Eq. (\ref{xi2}) for $E=0$ \cite{band1}.
Even when $E \ne 0$, we can evaluate the corresponding one period map $M(L)$ and its monodromy eigenvalues.
To adopt Aubry's band theory \cite{Aubry1,Aubry2},
we denote the argument of the eigenvalues of $M(L)$ by $-\pi \le \theta \le \pi$, rather than phase advance per period $\sigma \ge 0$.
Then, the distribution of points $(\theta, E)$ in the $\theta$-$E$ plane reveals a certain band structure (see Sec. \ref{VI} for numerical examples).
The dispersion curve $E(\theta)$ is symmetric with respect to $\theta = 0$,
and $dE(\theta)/d\theta = 0$ at $\theta = 0$ \cite{band1, band2}.
For the linear stability of the system,
there must exist 4 points (for 2 degrees of freedom, which is the case considered in this study) that
cross the $E = 0$ line.
As discussed in Ref. \cite{Aubry1}, the Krein signature is indeed the minus sign
of the slope of $E(\theta)$ at $E = 0$ for a given stable eigenvalue $\lambda = \exp( i \theta_0)$.
\begin{equation}
\kappa (\lambda) = - {\rm sgn} \left[ \frac{d E(\theta)}{d \theta}\bigg{|}_{\theta = \theta_0} \right],
\end{equation}
where $E(\theta_0) = 0$.

If a system parameter varies from a linearly stable condition, the band structure of $E(\theta)$ will evolve continuously.
Eventually, at some point, the dispersion curve may lose the intersection points with $E=0$.
This is indeed the onset of instability.
Just before the dispersion curve $E(\theta)$ becomes tangent to $E=0$ axis,
there are two intersection points $\theta_0 = \theta_l(\theta_r)$ on the left(right)-hand side of the tangential point.
Since the slope of the dispersion curve $d E(\theta) / d \theta$ at $\theta_l$ and $\theta_r$ should have opposite signs,
the tangential point can be interpreted as a collision between eigenvalues of different Krein signatures
(see more illustrative explanation in Refs \cite{Aubry1, Aubry2}).
Once a band structure loses contact points with the $E=0$ axis, a band gap is formed in which
monodromy eigenvalues are no longer on the unit circle and instabilities occur.
The band structure analysis is complementary to the Krein theory and
helpful for an improved understanding on the onset of instabilities.

\section{Numerical Examples} \label{VI}

We numerically calculated the eigenvalues and eigenvectors of a one-period map, and
associated properties such as Krein signatures and band structures.
In some cases, analytical solutions for the eigenvalues of the one-period map are available.
For example, in Refs. \cite{CS,Conte_MacKay}, the difference resonance and sum resonance of the single-particle motion in the ring
have been investigated in terms of approximate analytical expressions.
To establish a general theoretical framework that can be applied in various cases of single-particle motions and envelope perturbations,
we adopt {\rm MATHTEMATICA} \cite{Mathematica} that can handle any form of focusing matrix and one-period map (composed of either analytical expressions or numerical values).
The variables $s, x$, and $y$ are normalized by $L$.
The focusing strengths $\kappa_x, \kappa_y$, and $\kappa_{sq}$ are normalized by $1/L^2$.
The beam envelopes $a(s)$ and $b(s)$ are normalized by $\sqrt{\epsilon L}$ with $\epsilon_x = \epsilon_y = \epsilon$.
The self-field perveance $K_b$ is normalized by $\epsilon/L$.

\subsection{Single-particle motions with skew quadrupoles}\label{Sec_single_example}
\begin{figure}
  \centering
  \includegraphics[width=12cm]{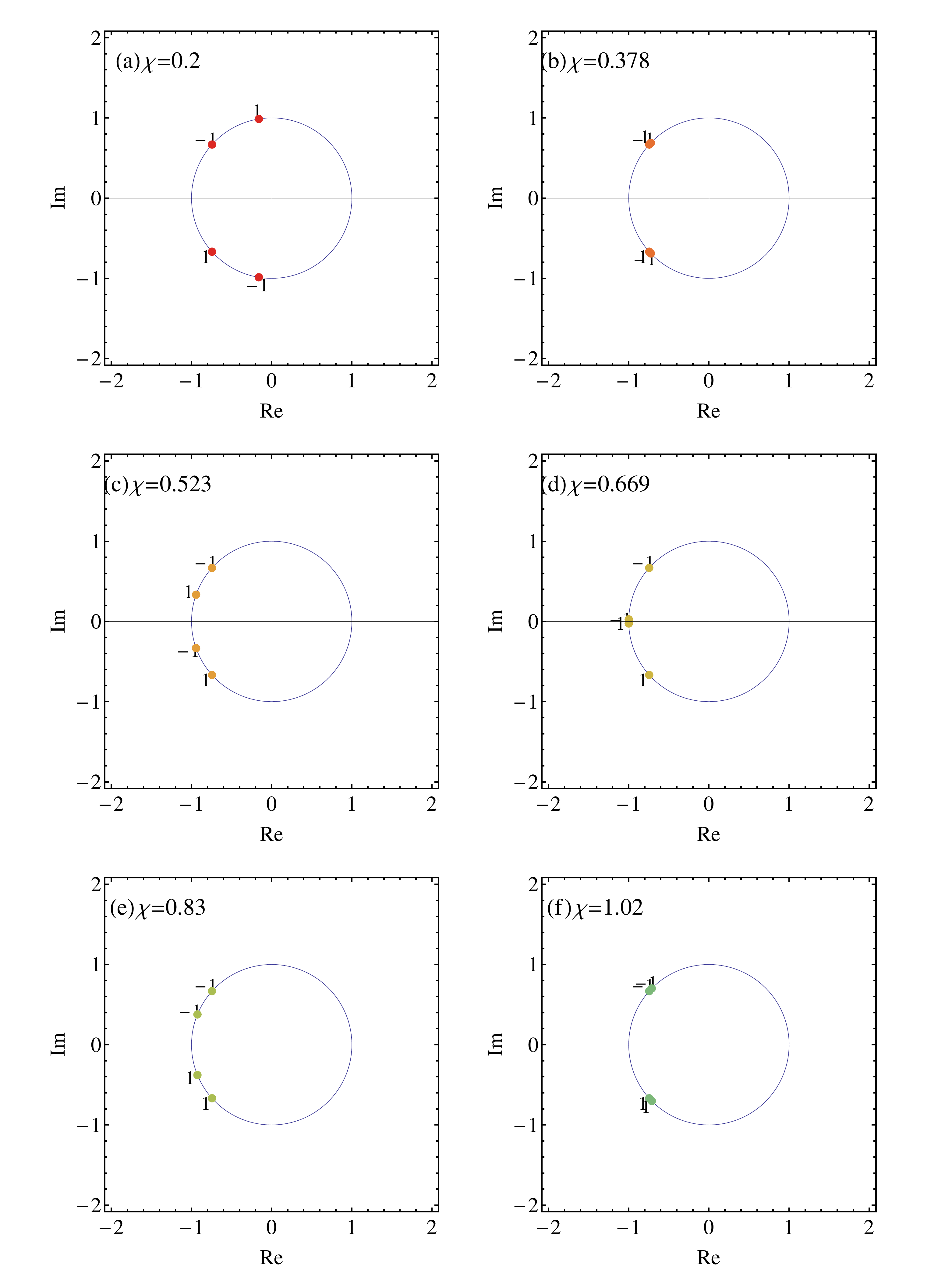}\\
  \caption{
  Evolution of the eigenvalues in the complex plane according to the changes in $\chi = \kappa_y / \kappa_x$.
  Here, $\kappa_{sf} = 15$ and $\hat{\kappa}_{sq} = 0$.
  The colors of the dots match those in Fig. \ref{Fig5} for corresponding $\chi$.
  See supplementary material for animation (Fig2animation.gif).
  }
  \label{Fig2}
\end{figure}

Figure \ref{Fig2} illustrates
the evolution of the eigenvalues in the complex plane with the increase of $\chi = \kappa_y / \kappa_x$.
The numbers ``$-1$'' and ``$1$'' next to the eigenvalues (represented by dots) indicate the corresponding Krein signatures.
Given that there is no coupling in this case ($\hat{\kappa}_{sq} = 0$),
the eigenvalues remain on the unit circle despite the fact that there are three collision points [Figs. \ref{Fig2}(b), \ref{Fig2}(d), and \ref{Fig2}(f)].
They simply bypass each other.
However, this does not mean that all three collision points are strongly stable.

\begin{figure}
  \centering
  \includegraphics[width=12cm]{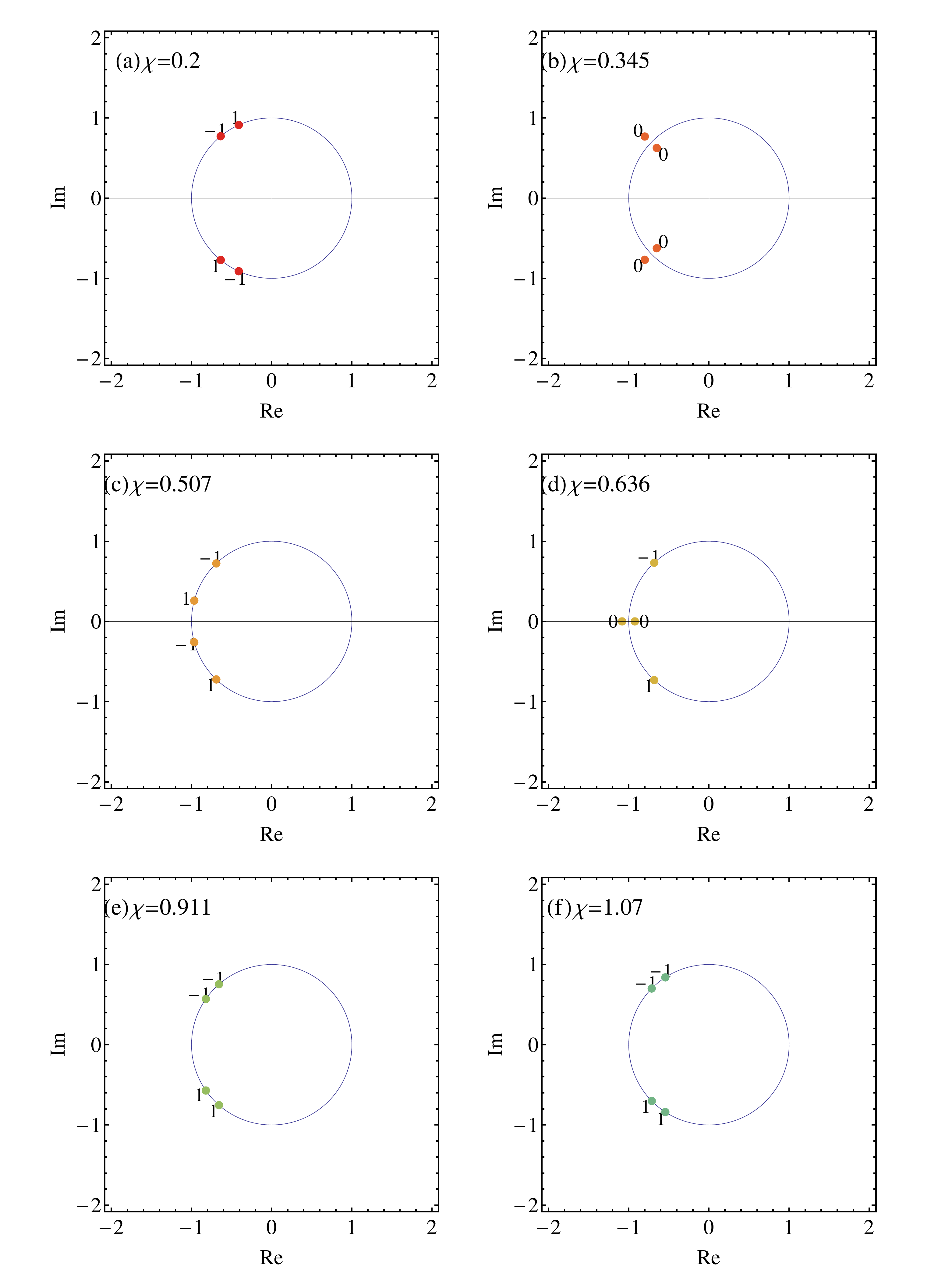}\\
  \caption{
  Evolution of the eigenvalues in the complex plane according to the changes in $\chi = \kappa_y / \kappa_x$.
  Here, $\kappa_{sf} = 15$, and $\hat{\kappa}_{sq} = 5$ with $\eta = 0.3$ and $\alpha = 0.5$.
  The colors of the dots with those in Fig. \ref{Fig5} for corresponding $\chi$.
  See supplementary material for animation (Fig3animation.gif).
  }
  \label{Fig3}
\end{figure}

If we introduce a coupling, then eventually instabilities occur as shown in Fig. \ref{Fig3}.
Figure \ref{Fig3}(b) corresponds to the sum resonance, which is
caused by the collision between eigenvalues of positive and negative Krein signatures.
Because we fix $\kappa_x = \kappa_{sf}$ and increase $\kappa_y$ by $\chi \kappa_{sf}$,
we denote the phase advance of the eigenvalue of positive signature moving counterclockwise by $\sigma_{2}$,
and that of the stationary eigenvalue of positive signature by $\sigma_{1}$.
The condition for the sum resonance is then $\sigma_{1} + \sigma_{2} = 360^\circ$, as expected.
Note that as the coupling strength increases, the eigenmode phase advance $\sigma_{1}$($\sigma_{2}$) deviates from $\sigma_{0x}$($\sigma_{0y}$),
which is the phase advance in the $x(y)$-direction in the absence of coupling.

Figure \ref{Fig3}(d) corresponds to the so-called half-integer resonance, which is also associated
with the Krein collision of opposite signatures.
After the collision at $\sigma_{2} = 180^\circ$, the eigenvalues move off the unit circle.

By contrast, the same signature collision reveals a very different feature, as shown in Figs. \ref{Fig3}(e) and \ref{Fig3}(f).
As one eigenvalue approaches to a stationary one, they repel each other rather than overlapping.
Eventually, the initially rotating eigenvalue becomes stationary, and the stationary eigenvalue starts rotating.
Indeed, both modes are exchanging their roles.
This process occurs around $\sigma_{1} - \sigma_{2} = 0$ or $\chi = 1$, and has been known as difference resonance.
The difference resonance is not as dangerous as the sum resonance, but it enhances an exchange of betatron oscillations
between horizontal ($x$) and vertical ($y$) planes \cite{Wolski.book}.

\begin{figure}
\centering
\subfigure[$\hat{\kappa}_{sq}=0$.]
{
\includegraphics[height=0.2 \textheight]{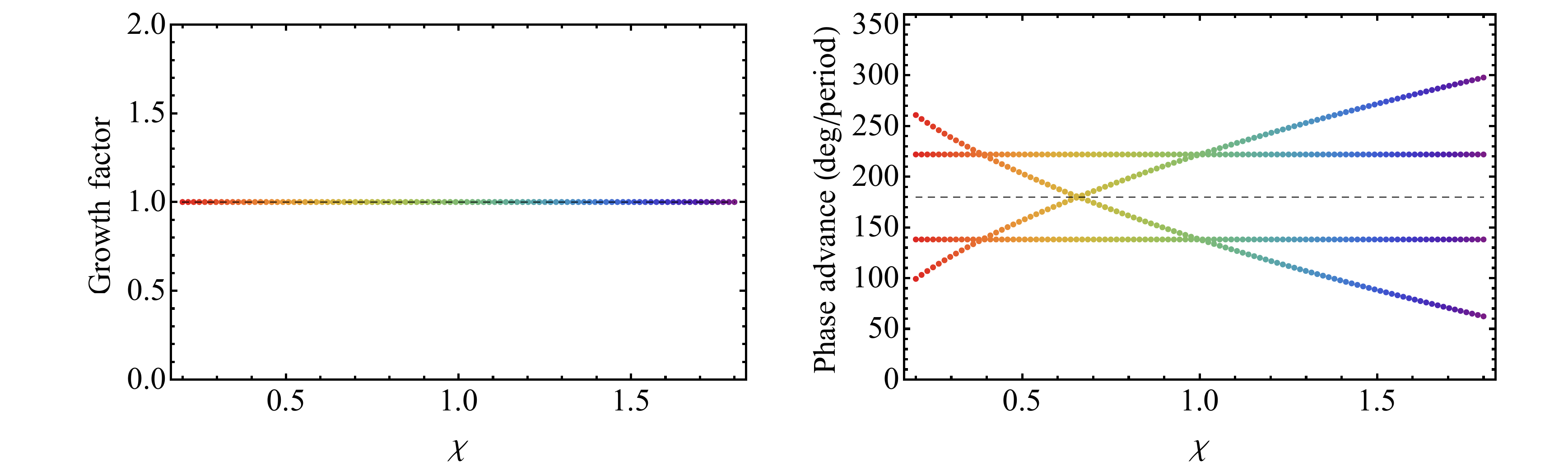}
}
\vspace*{5mm}
\\
\subfigure[$\hat{\kappa}_{sq}=3$.]
{
\includegraphics[height=0.2 \textheight]{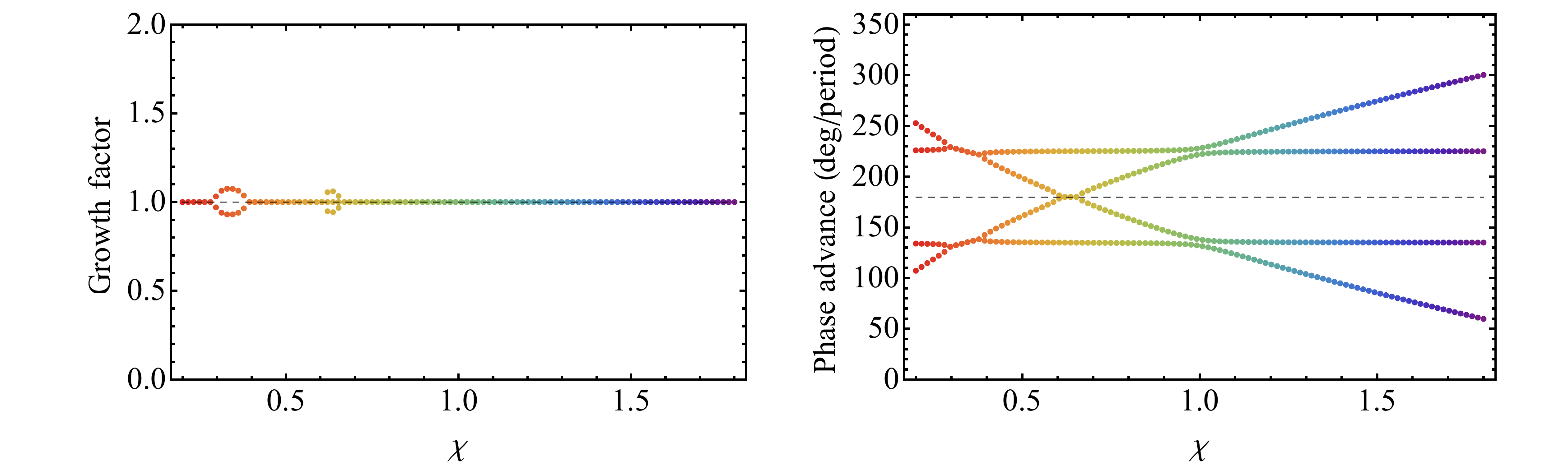}
}
\vspace*{5mm}
\\
\subfigure[$\hat{\kappa}_{sq}=5$.]
{
\includegraphics[height=0.2 \textheight]{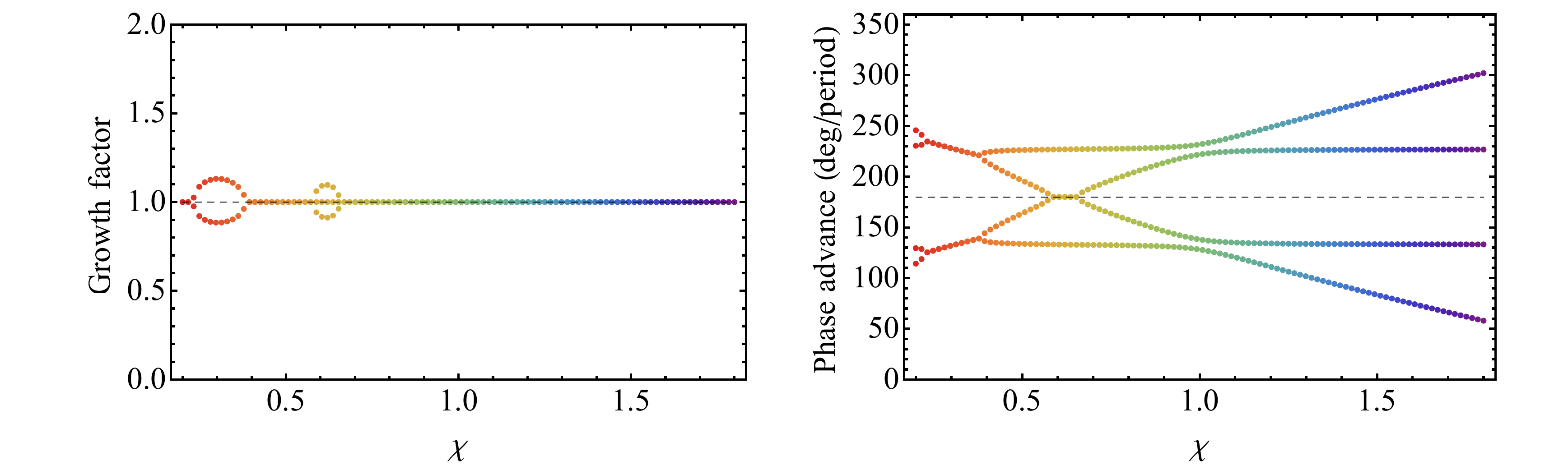}
}
\caption[]
{
Growth factors (left) and phase advances per period (right) of the single-particle motions
as functions of the asymmetric focusing factor $\chi = \kappa_y / \kappa_x$
for several different coupling strengths $\hat{\kappa}_{sq}$.
}
\label{Fig4}
\end{figure}

The growth factors ($\gamma_n$) and phase advances ($\sigma_n$) per period of the eigenmodes are plotted in Fig. \ref{Fig4}
for several different coupling strengths $\hat{\kappa}_{sq}$.
It is clearly seen that the growth factors and instability bands are increasing according to $\hat{\kappa}_{sq}$.
Moreover, the separation gap of two eigenmodes near $\chi = 1$ (or, $\sigma_{1} = \sigma_{2}$) is widened with the increase
of the coupling strength $\hat{\kappa}_{sq}$.

\begin{figure}
\centering
\subfigure[$\hat{\kappa}_{sq} = 0$ and $\chi=0.387$ with $\sigma_{0x}=221.9^\circ$ and $\sigma_{0y}=138.1^\circ$.]
{
\includegraphics[height=0.25 \textheight]{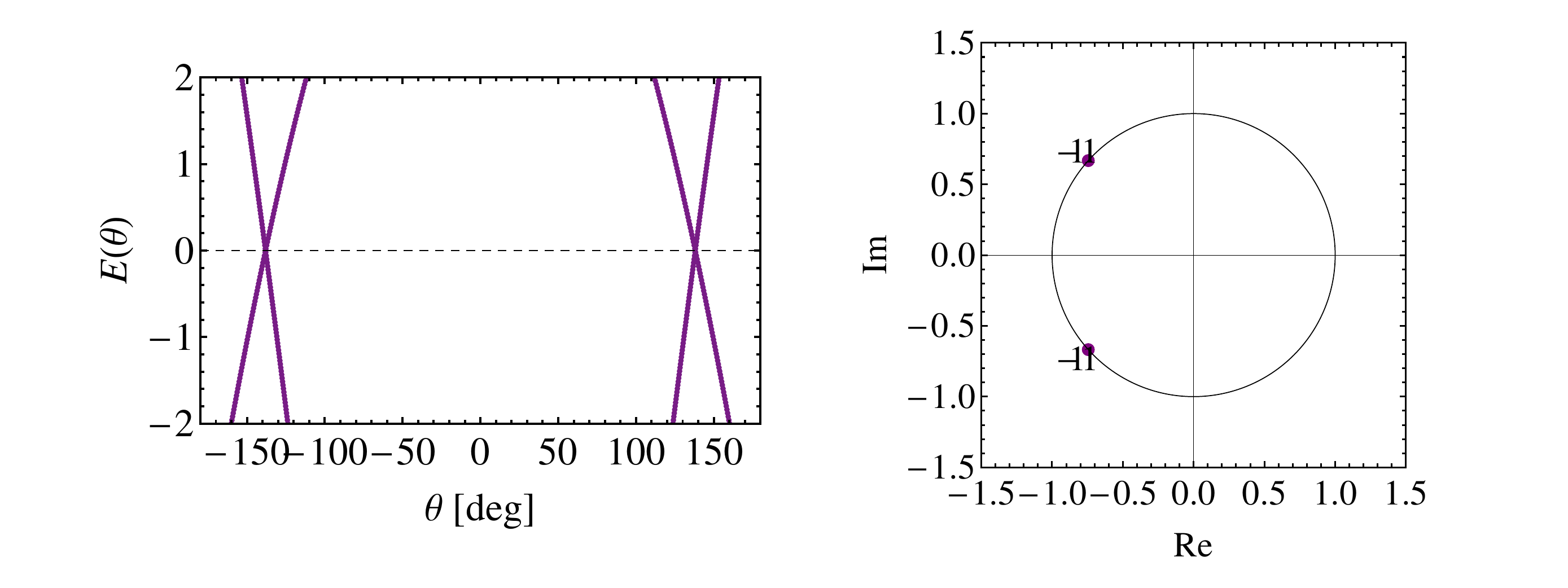}
}
\\
\subfigure[$\hat{\kappa}_{sq} = 3$ and $\chi=0.387$ with $\sigma_{0x}=221.9^\circ$ and $\sigma_{0y}=138.1^\circ$.]
{
\includegraphics[height=0.25 \textheight]{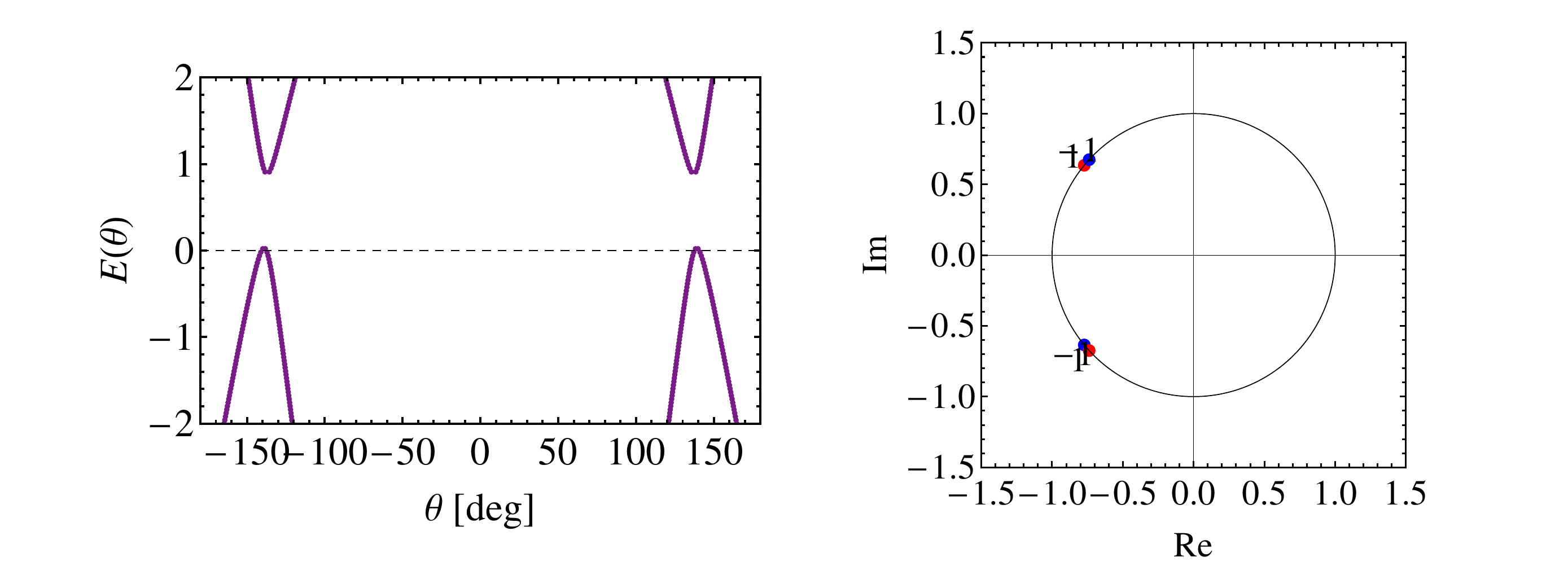}
}
\\
\subfigure[$\hat{\kappa}_{sq} = 3$ and $\chi=0.349$ with $\sigma_{0x}=221.9^\circ$ and $\sigma_{0y}=131.0^\circ$.]
{
\includegraphics[height=0.25 \textheight]{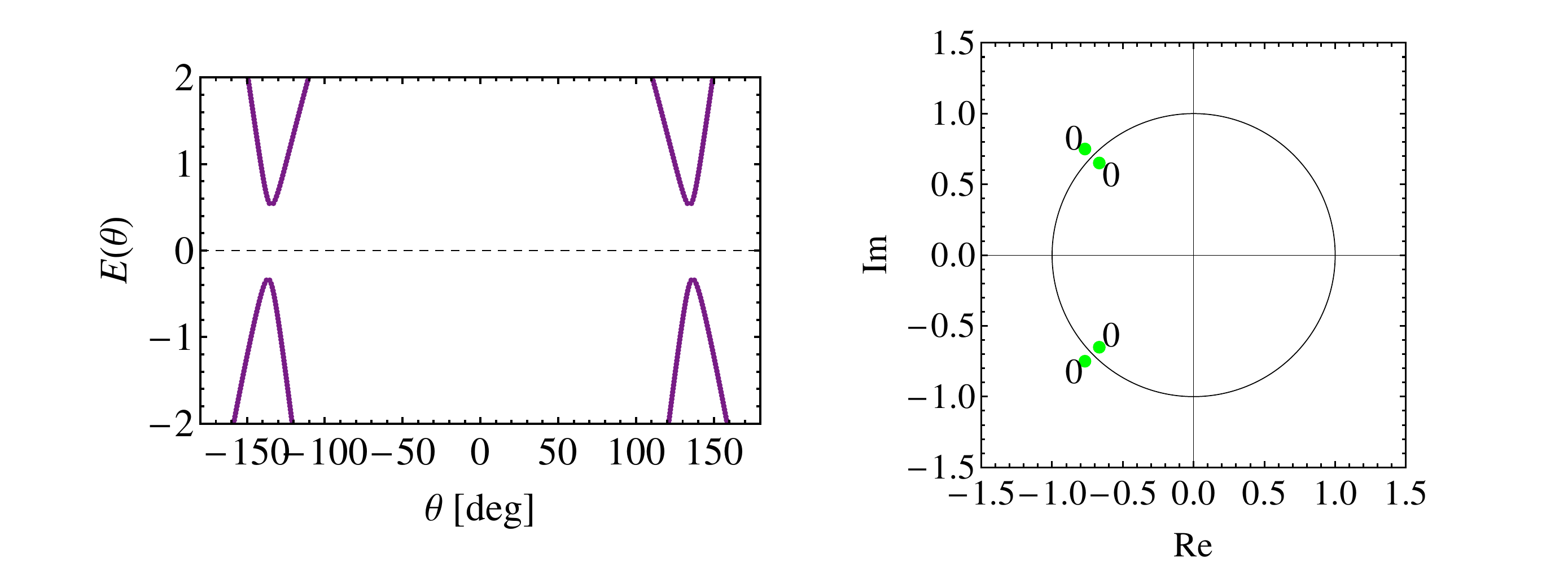}
}
\caption[]
{
Band structures (left) and eigenvalues (right) near the sum resonances.
Here, $\sigma_{0x}(\sigma_{0y})$ is the phase advance per period of the single particle motion in the $x(y)$-direction
in the absence of coupling.
}
\label{Fig5}
\end{figure}

To get further insight on the sum resonance, we apply band analysis.
Figure \ref{Fig5}(a) shows the band structure at the sum resonance for an uncoupled case.
The upper and lower bands are connected, and the dispersion curve still maintains its contacts with $E=0$ line, indicating
no immediate appearance of the instability, as expected.
When we apply a certain level of coupling, a band gap is formed as demonstrated in Fig. \ref{Fig5}(b).
Given that the phase advances are tuned for $\sigma_{0x} + \sigma_{0y} = 360^\circ$  but not for $\sigma_{1} + \sigma_{2} = 360^\circ$,
the dispersion curve is still tangential to the $E=0$ line.
If we slightly lower $\chi$ so that $\sigma_{1} + \sigma_{2} = 360^\circ$ is maintained,
we then shift the band gap lying around $E=0$ line and finally observe an instability [see Fig. \ref{Fig5}(c)].

\subsection{Envelope perturbations in high-intensity beams} \label{Sec_envelope_example}

\begin{figure}
\centering
\subfigure[$\hat{\kappa}_q = 15, \eta = 0.3, \alpha = 0.5$, and $\chi = 1$ with $\sigma_{vx} = 60.4^\circ$ and $\sigma_{vy} = 60.4^\circ$]
{
\includegraphics[height=0.2 \textheight]{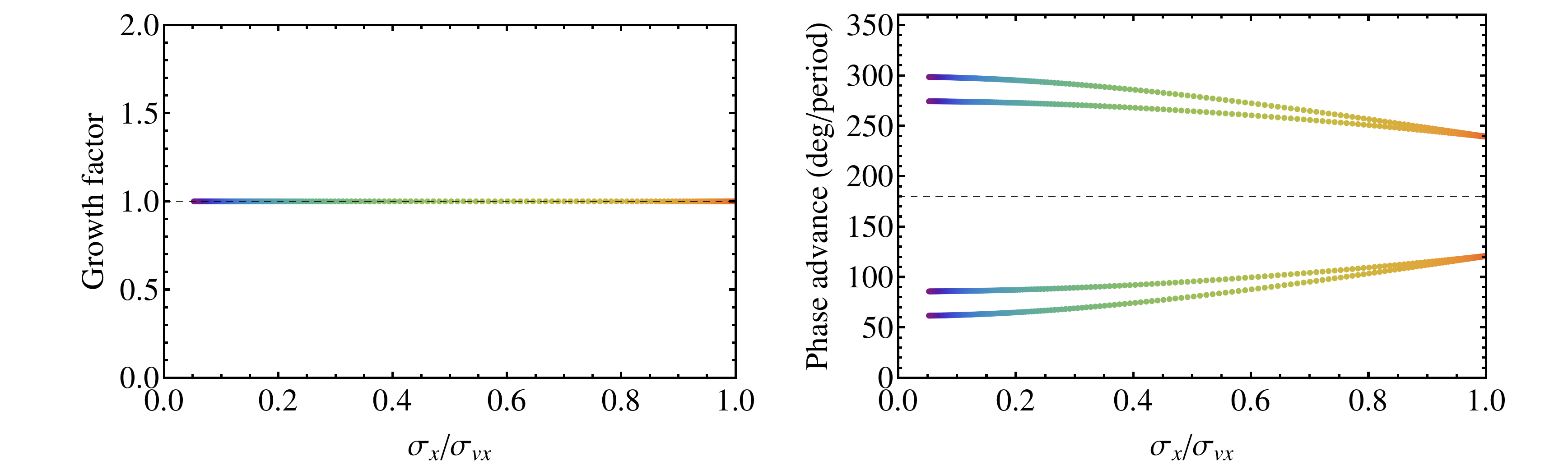}
}
\vspace*{5mm}
\\
\subfigure[$\hat{\kappa}_q = 26, \eta = 0.3, \alpha = 0.5$, and $\chi = 1$ with $\sigma_{vx} = 121.1^\circ$ and $\sigma_{vy} = 121.1^\circ$]
{
\includegraphics[height=0.2 \textheight]{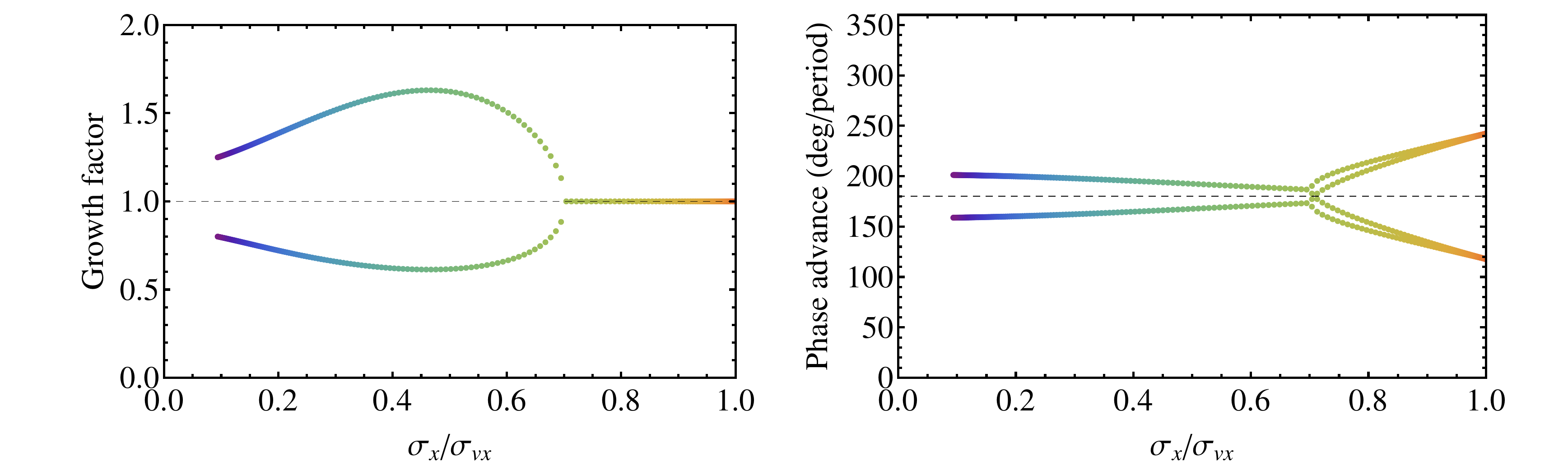}
}
\vspace*{5mm}
\\
\subfigure[$\hat{\kappa}_q = 26, \eta = 0.3, \alpha = 0.5$, and $\chi = 0.97$ with $\sigma_{vx} = 121.7^\circ$ and $\sigma_{vy} = 114.6^\circ$]
{
\includegraphics[height=0.2 \textheight]{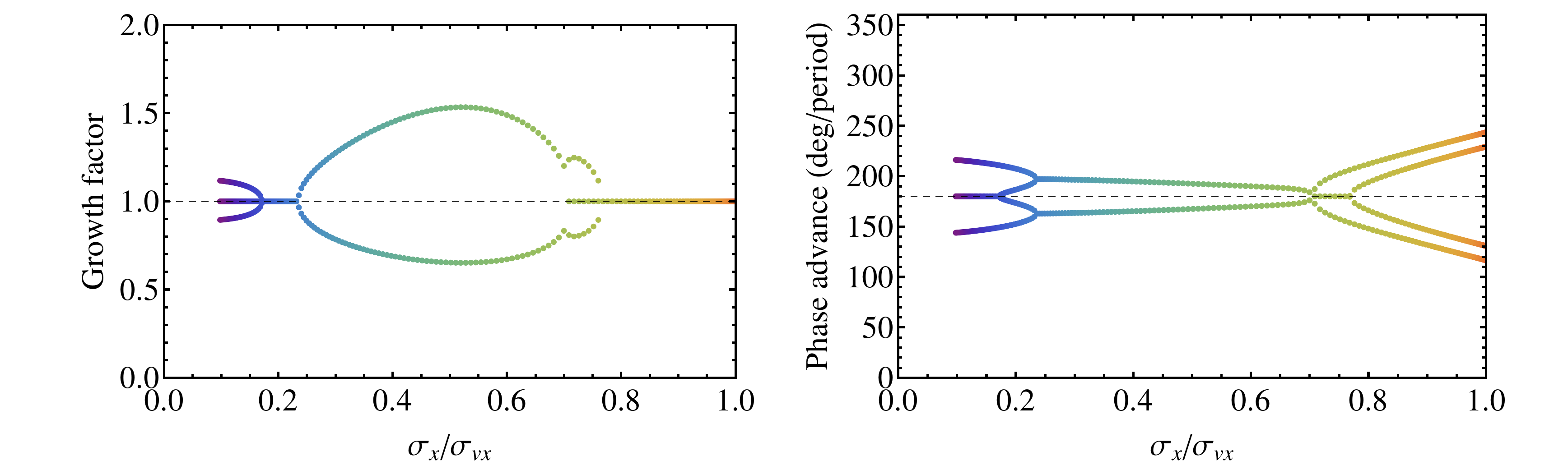}
}
\caption[]
{
Growth factors (left) and phase advances per period (right) of the envelope perturbations
as functions of the space-charge tune depression $\sigma_x / \sigma_{vx}$
for several different lattice configurations.
Here, $\sigma_{vx}$ is the phase advance per period of the single particle motion in the $x$-direction
in the absence of space charge ($K_b =0$).
}
\label{Fig6}
\end{figure}

The growth factors ($\gamma_n$) and phase advances ($\sigma_n$) per period of the envelope perturbations
as functions of the space-charge tune depression $\sigma_x / \sigma_{vx}$
are plotted in Fig. \ref{Fig6} for several different lattice settings.
Here, $\sigma_{x}$ is the single-particle phase advance in the $x$-direction
and $\sigma_{vx}$ is without space charge ($K_b = 0$).
It should be emphasized here that the phase advances for envelope oscillation eigenmodes ($\sigma_{1}$ and $\sigma_{2}$) and those for single particle orbits ($\sigma_x$ and $\sigma_y$) are generally different.
In the limit of zero space charge, it is well-known that
the phase advance of $\delta a$ ($\delta b$) oscillations is twice the single-particle vacuum phase advance $\sigma_{vx}$ ($\sigma_{vy}$) \cite{Struckmeier, Lund.PRAB2004}.
The lager phase advance is assigned to $\sigma_1$ whereas the smaller is assigned to $\sigma_2$.
The eigenmodes associated with $\sigma_1$ and $\sigma_2$ have
analogous characteristics to breathing and quadrupole modes in a smooth-focusing channel, respectively \cite{Lund.PRAB2004}.

In Fig. \ref{Fig6}(a), we set $\sigma_{vx} = \sigma_{vy} = 60.4^\circ$.
Therefore, the phase advances of two positive-signature eigenmodes start from $\sigma_{1} = \sigma_{2} = 120.8^\circ$ at  $\sigma_x / \sigma_{vx} =1$.
Note that the space-charge tune depression $\sigma_x / \sigma_{vx}$ is decreasing from 1
as we increase the self-field perveance $K_b$ from 0.
At first glance, it appears that as $K_b$ increases, the $k_{0m}$ term in Eq. (\ref{km}) increases accordingly,
and thus average the focusing effect on
the envelope perturbations becomes stronger.
However, this is not the case. Because the matched beam envelopes $a_m$ and $b_m$ increase at the same time, the overall effect of increase in $K_b$ turns out
to be a reduction in the average focusing.
Therefore, we observe that $\sigma_{1}$ and $\sigma_{2}$ decreases from their initial value of $120.8^\circ$
as $\sigma_x / \sigma_{vx}$ decreases from 1.

\begin{figure}
  \centering
  \includegraphics[width=12cm]{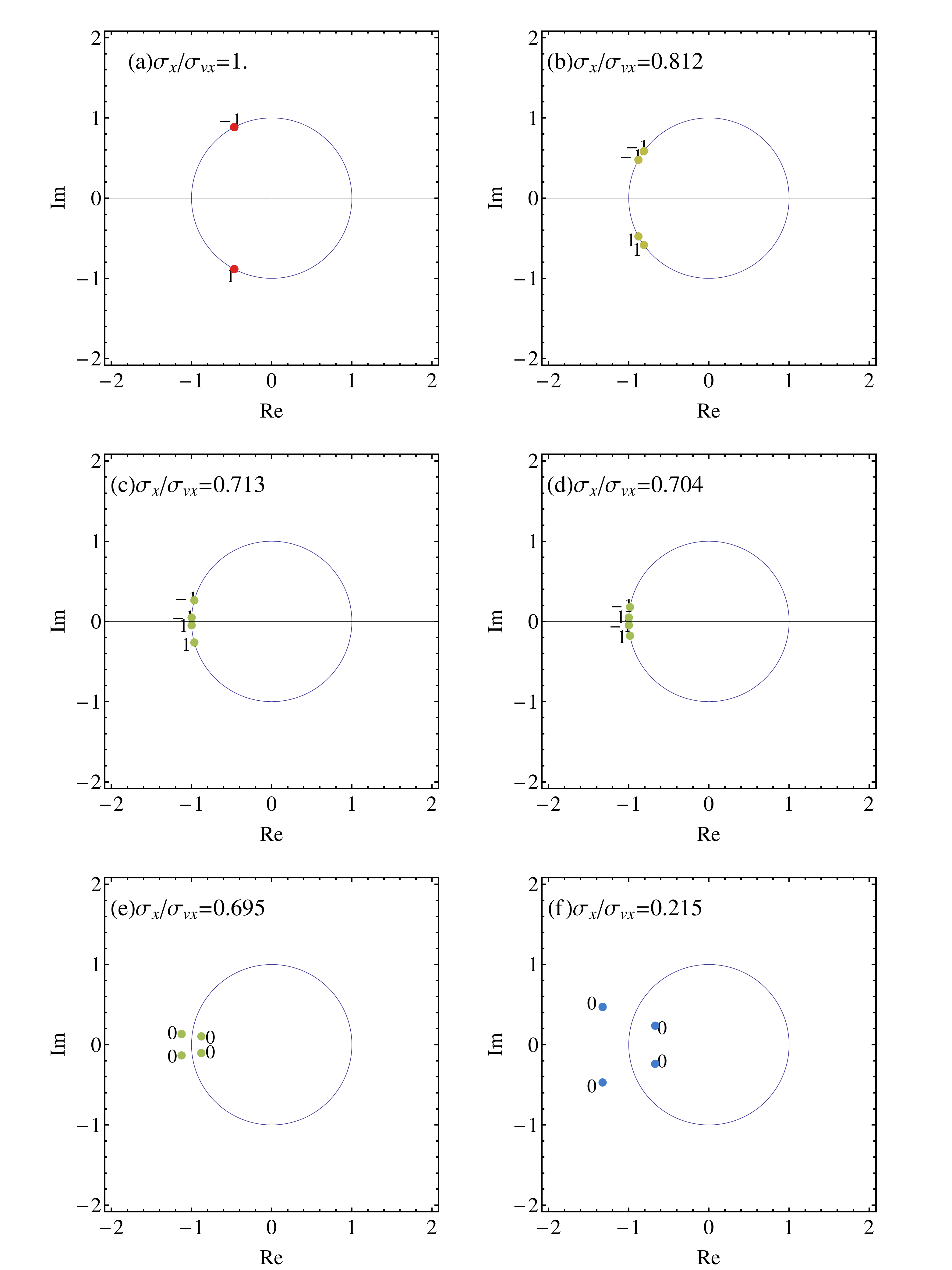}\\
  \caption{
  Evolution of the eigenvalues in the complex plane according to the changes in $K_b$, or equivalently $\sigma_x / \sigma_{vx}$.
  Here, $\hat{\kappa}_q = 26, \eta = 0.3, \alpha = 0.5$, and $\chi = 1$.
  The colors of the dots match those in Fig. \ref{Fig6} for corresponding $\sigma_x / \sigma_{vx}$.
  See supplementary material for animation (Fig7animation.gif).
  }
  \label{Fig7}
\end{figure}

By contrast, if we set $\sigma_{vx}$ and $\sigma_{vy}$ above $90^\circ$,
$\sigma_{1}$ and $\sigma_{2}$ of positive Krein signature modes start above $180^\circ$ [see Figs. \ref{Fig6}(b) and \ref{Fig6}(c)].
As $\sigma_x / \sigma_{vx}$ decreases, the eigenvalues approach to $180^\circ$ line.
In Fig. \ref{Fig6}(b), the first Krein collision at $180^\circ$ does not produce an instability.
Instead, the eigenvalues pass through each other and
make collisions with other eigenvalues following behind [see Figs. \ref{Fig7}(c) and \ref{Fig7}(d)].
Eventually, eigenvalues move off the unit circle and trigger an instability [see Figs. \ref{Fig7}(e) and \ref{Fig7}(f)].
Two phase advances $\sigma_1$ and $\sigma_2$ are locked to $\sigma_1 + \sigma_2 = 360^\circ$,
with one above $180^\circ$ and the other below $180^\circ$, respectively.
This parametric instability is often called a confluent resonance because
it involves both envelope oscillation mode frequencies \cite{Reiser.book,Lund.PRAB2004}.
Indeed, the confluent resonance is mathematically equivalent to the sum resonance of single-particle motion.
Both resonances are the outcome of the collision between
eigenvalues of positive and negative Krein signatures, and their phase advances are locked to $\sigma_1 + \sigma_2 = 360^\circ$.
It can be interpreted from a physical point of view as follows:
given that the total action has to be conserved for the linear system \cite{Bridges, action},
an unstable mode should have equal amounts of positive and negative actions \cite{accretion}.

If we slightly lower $\chi$  so that $\sigma_{vx} \ne \sigma_{vy}$ and the separation between $\sigma_1$ and $\sigma_2$ becomes larger,
then we observe the clear appearance of a lattice resonance at $\sigma_2 = 180^\circ$ [see Fig. \ref{Fig6}(c)].
The lattice resonance is a type of parametric instability that
represents a half-integer resonance between the focusing structure and one of the mode oscillation frequencies \cite{Lund.PRAB2004}.
Mathematically, this lattice resonance is analogous to the half-integer resonance of single-particle motion.

\begin{figure}
\centering
\subfigure[$K_b = 1.38$, $\sigma_x / \sigma_{vx}= 0.71$,  and $\chi = 1$.]
{
\includegraphics[height=0.25 \textheight]{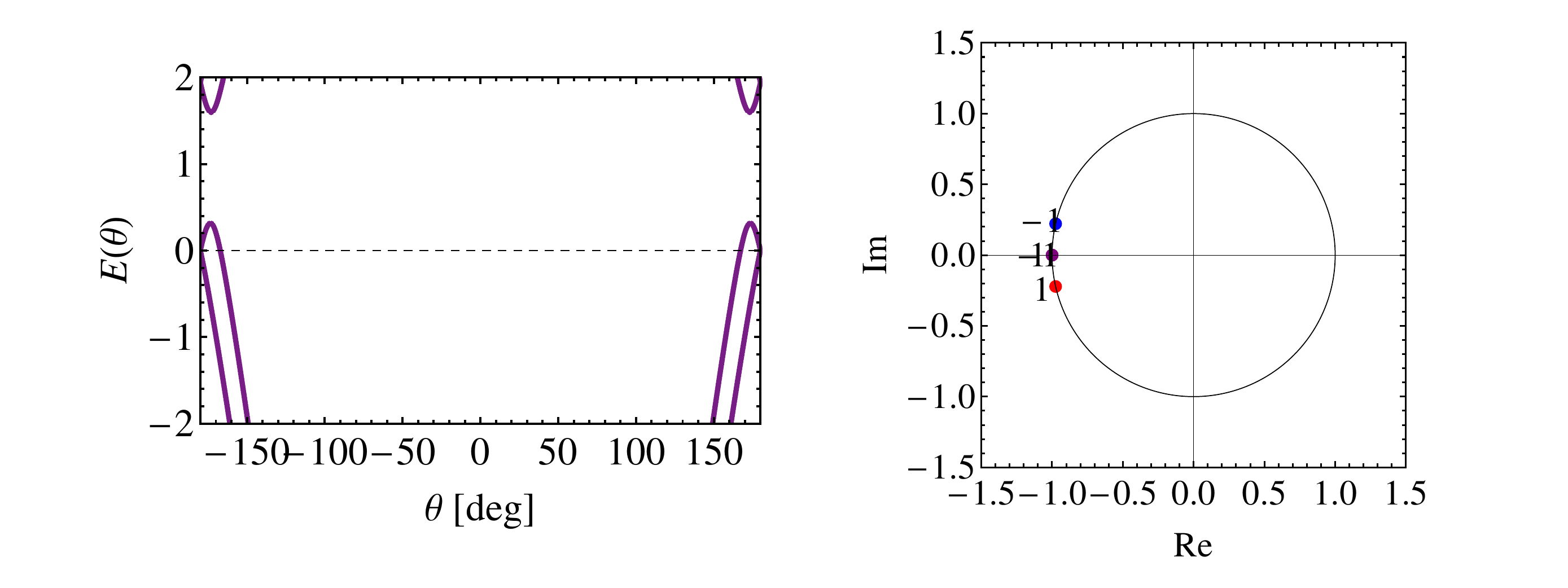}
}
\\
\subfigure[$K_b = 1.38$, $\sigma_x / \sigma_{vx}= 0.71$,  and $\chi = 0.99$.]
{
\includegraphics[height=0.25 \textheight]{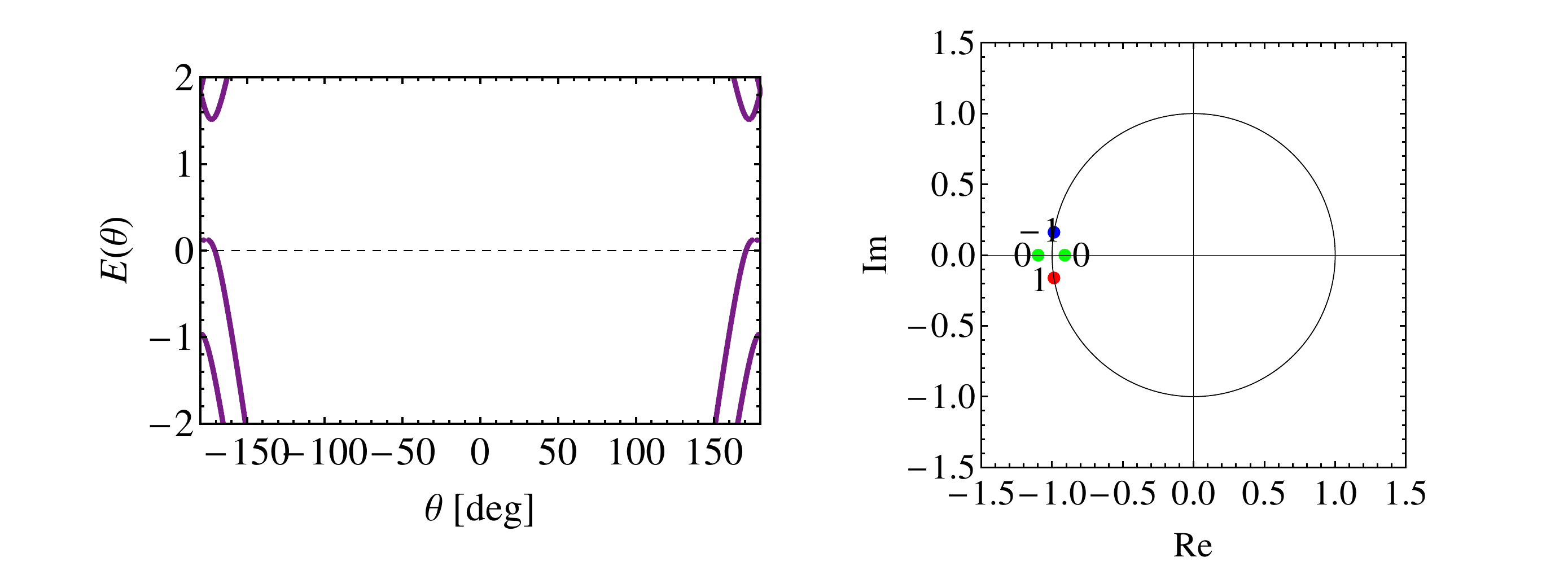}
}
\\
\subfigure[$K_b = 1.38$, $\sigma_x / \sigma_{vx}= 0.71$,  and $\chi = 1.01$.]
{
\includegraphics[height=0.25 \textheight]{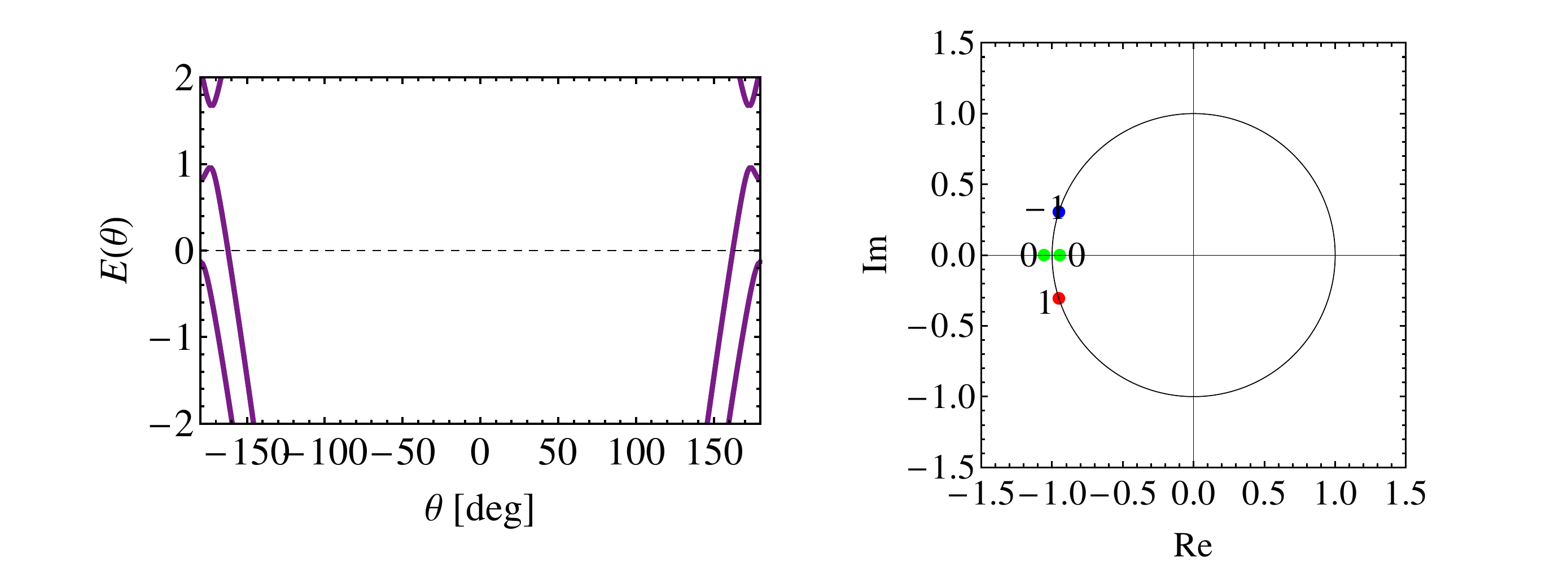}
}
\caption[]
{
Band structures (left) and eigenvalues (right) near the lattice resonances of the envelope perturbations.
}
\label{Fig8}
\end{figure}

To investigate why there was no lattice resonance in Figs. \ref{Fig6}(b), \ref{Fig7}(c), and \ref{Fig7}(d),
we apply band structure analysis.
Figure \ref{Fig8}(a) shows the band structure at $\sigma_x / \sigma_{vx} =0.71$ in which $\sigma_2 = 180^\circ$.
The dispersion curve still maintains its contacts with $E=0$ line at $\theta = \pm 180^\circ$ despite the Krein collision.
We changed system parameters such as $\hat{\kappa}_q, \eta$, and $\alpha$ continuously.
However, we could not
observe the formation of a band gap at $\theta = \pm 180^\circ$.
Instead, changes in the asymmetric focusing factor $\chi$ induce a breakup of the dispersion curve and generate a band gap at $\theta = \pm 180^\circ$
as depicted in Figs. \ref{Fig8}(b) and \ref{Fig8}(c).
Therefore, the Krein collision of opposite signatures at $\sigma_2 = 180^\circ$ is indeed not strongly stable,
again manifesting the Krein-Gel’fand-Lidskii theorem.

\section{Conclusions and Future Work} \label{VII}
We revisited the general problem of linear beam stability in periodic focusing systems in a pedagogical aspect.
In particular, we apply recent understandings on the Krein theory and band structure analysis to
two of the most fundamental stability problems in accelerator physics, namely
single-particle motions with skew quadrupoles and envelope perturbations in high-intensity beams.
We clarified through numerical examples that
the physical meaning of the Krein signature is the sign of the action for the eigenmode,
and the only path for instability is via the resonance between the positive- and  negative-action modes.
By monitoring the Krein collisions together with the signatures of the involved eigenvalues,
one can easily explore the parameter space for instabilities.
For example, we identified that the sum resonance in single-particle motion
is mathematically equivalent to the confluent resonance in the envelope oscillation
in the sense that both parametric instabilities are the outcome of the Krein collision of opposite signatures at
phase advances other than $180^\circ$.
The band structure analysis is complementary to the Krein theory and turns out to be
useful for a detailed understanding on the appearance of the instabilities.

Next, some of the future scope are also briefly discussed.
While we provide only examples with two degrees of freedom,
all the properties of the Krein collision discussed in this study are readily applicable
to cases with three or higher degrees of freedom.

For non-homogeneous systems, i.e., ${\bf z}' = K(s) {\bf z} + {\bf f}(s)$,
only limited studies have been carried out on the stability analysis within the context of the Floquet theory \cite{Non-homogeneous1}.
The external forcing vector ${\bf f}(s)$ can induce resonances when the driving frequency matches
the frequency of the periodic solution of the homogenous equation ${\bf z}' = K(s) {\bf z}$ \cite{Non-homogeneous2}.
For the driven oscillations, the amplitudes of the solutions grow linearly on resonances.
By contrast, for the parametric instabilities in homogeneous systems, the solutions typically have exponential growths.
We may investigate new resonance lines in the parameter space
after properly constructing the forcing vector.

A complex generalization of the Krein collision in terms of the linear G-Hamiltonian system can be found in Refs. \cite{Yakubovich.book, Zhang.PoP}.
For some instability problems, complex coefficients may appear in the focusing matrix $K(s)$
after Fourier decomposition of the perturbations $g(s, t)$ in the form of $ \tilde{g}(k, \omega) \exp(i k s - i\omega t) $.
In these situations,
stability analysis based on the G-Hamiltonian structure and Krein collision should be an effective framework.

Finally, we emphasize the importance of the phase advances in one-period map $M(L)$.
It is well established in Refs. \cite{GCS.PRSTAB,Spectral_Structural}
that a one-period map has an elegant parametrization for a stable system as follows:
$M(L)= Q_0^{-1} P(L)^{-1} Q_0$.
Here, $Q_0$ is an envelope matrix that allows canonical transformation into
a normalized coordinate system, and $P(L)^{-1}$ is the phase advance matrix that performs a symplectic rotation
in that normalized coordinates.
If $\psi$ is an eigenvector of $M(L)$, then $Q_0 \psi$ is the eigenvector of $P(L)^{-1}$.
Moreover, the eigenvalues (or phase advances) and Krein signatures of $P(L)^{-1}$ are identical to those of $M(L)$.
Hence, the stability properties can be completely determined by the phase advance matrix.
We note that $Q_0$ is unique up to a symplectic rotation, and thus we can always make $P(L)^{-1}$ to represent
a simultaneous rotation with given phase advances in each eigenplane.
A more detailed discussion on this aspect will be reported in the near future.

\section*{ACKNOWLEDGMENTS}
This research was supported by the National Research Foundation of Korea(Grant No. NRF-2017M1A7A1A02016413).
This work was also supported by the U.S. Department of Energy Grant No. DE-AC02-09CH11466.

\end{document}